\documentclass[12pt,reqno]{amsart}

\usepackage{amsmath,amsfonts,amsthm,amssymb}

\usepackage[
height=20cm,top=24mm,bottom=24mm,left=26mm,right=26mm]{geometry}

\usepackage[numbers,sort]{natbib}

\usepackage[]{graphicx}
\numberwithin{equation}{section}
\allowdisplaybreaks[1]

\newcommand{\I}{\mathrm{i}}
\newcommand{\E}{\mathrm{e}}

\DeclareMathOperator{\CS}{CS}

\DeclareMathOperator{\sign}{sgn}
\DeclareMathOperator{\Tr}{Tr}

\DeclareMathOperator{\ch}{ch}
\DeclareMathOperator{\erfc}{erfc}

\DeclareMathDelimiter{\Norm}{\mathord}{largesymbols}{"3E}{largesymbols}{"3E}

\begin{document}
\baselineskip 16pt
\parskip 8pt
\sloppy


\title[Superconformal Algebras and Mock Theta Functions 2]{
  Superconformal Algebras and Mock Theta Functions 2. Rademacher
  expansion for K3 Surface
}



\author[T. Eguchi]{Tohru \textsc{Eguchi}}

\author[K. Hikami]{Kazuhiro \textsc{Hikami}}


\address{Yukawa Institute for Theoretical Physics, Kyoto University,
  Kyoto 606--8502, Japan}
\email{
  \texttt{eguchi@yukawa.kyoto-u.ac.jp}
}

\address{Department of Mathematics, 
  Naruto University of Education,
  Tokushima 772-8502, Japan.}

\email{
  \texttt{KHikami@gmail.com}
}


\date{April 6, 2009.
  Revised on September  14, 2009.
}

\begin{abstract}
The elliptic genera of the K3 surfaces, both compact and non-compact
cases,
are studied by using the theory of mock
theta functions.
We decompose the elliptic genus in terms of the $\mathcal{N}=4$
superconformal characters at level-$1$, and
present an
exact formula for the coefficients of the massive (non-BPS) representations
using  Poincar{\'e}--Maass series.
\end{abstract}




\maketitle

\section{Introduction}

Studies
of an asymptotic behaviour of $p(n)$, which is the number of
partitions of $n$ defined by
\begin{equation*}
  \prod_{n=1}^\infty
  \frac{1}{1-q^n}
  =
  \sum_{n=0}^\infty p(n) \, q^n ,
\end{equation*}
were initiated
by Hardy and Ramanujan~\cite{HardyRaman18a}.
The generating function in the left hand side
is essentially the inverse of the Dedekind $\eta$ function, and 
it is well-known that
its asymptotic behavior is given by
\begin{equation*}
  p(n) \sim
  \frac{1}{4 \, n \, \sqrt{3} } \,
  \E^{\pi \,
    \sqrt{\frac{2 \, n}{3}}
  } .
\end{equation*}
An exact asymptotic expansion was later derived by Rademacher by use of
the circle method
(\emph{e.g.}~\citep[Chapter 14]{HRadema73}) as
\begin{equation*}
  p(n) =
  \frac{\pi}{
    \left( 24 \, n - 1 \right)^{\frac{3}{4}}
  } \,
  \sum_{c=1}^\infty
  \frac{1}{ \sqrt{12 \, c}} \,
  I_{\frac{3}{2}} \left(
    \frac{\pi}{6 \,c} \,
    \sqrt{24 \,n -1}
  \right)
  \sum_{\substack{
      d \mod 24 \, c \\
      d^2 =
      -24 \, n+1
      \mod 24 \, c
    }}
  \left(
    \frac{12}{d}
  \right) \,
  \E^{\frac{d}{6 \, c} \, \pi \, \I} ,
\end{equation*}
Here ${I_{3/2}}$ denotes the modified Bessel-function and we have
introduced the Legendre symbol
$\left(\frac{12}{\bullet}\right)$.

The number of partition $p(n)$ has received wide
interests~\cite{Andre76}, and has played an important role in various
areas of mathematics and physics.
One of  the generalizations of $p(n)$ is given by 
the Ramanujan mock theta function~\cite{FJDyson44}
(see also Refs.~\citenum{GEAndre89a,GordMcIn09a} for a review),
\begin{align*}
  f(q) 
  & =
  1+
  \sum_{n=1}^\infty
  \frac{
    q^{n^2}
  }{
    (1+q)^2 \, (1+q^2 )^2
    \cdots
    (1+q^n )^2
  }
  \\
  & =
  1+ \sum_{n=1}^\infty
  \alpha(n) \, q^n .
\end{align*}
Asymptotic behavior  of $\alpha(n)$
was written   in Ramanujan's last
letter to Hardy as 
\begin{equation*}
  \alpha(n)
  \sim
  \frac{
    (-1)^{n-1}}{
    2 \, \sqrt{ n - \frac{1}{24}}
  } \,
  \E^{\pi \,
    \sqrt{\frac{n}{6} - \frac{1}{144}}
  },
\end{equation*}
which
was proved by Dragonette~\cite{Drago52a}.
This  asymptotics 
was later improved by Andrews~\cite{GEAndre66d}, and
an exact formula for $\alpha(n)$ called
the Andrews--Dragonette
identity was conjectured.
Bringmann and Ono proved the conjecture
using the work of Zwegers~\cite{BrinKOno06a} on mock theta functions.

In our previous paper~\cite{EguchiHikami08a},
we have shown that 
the theory of  mock theta  functions is useful in  studying
the elliptic genera of hyperK{\"a}hler manifolds in terms of the 
representations of 
$\mathcal{N}=4$ superconformal algebra.
We have used the results of Zwegers~\cite{Zweg02Thesis} that   mock
theta function is a holomorphic part of the harmonic Maass form with
weight-$1/2$
and
has a  weight-$3/2$ (vector) modular form as its ``shadow''
(see \emph{e.g.} Refs.~\citenum{Zagier08a,KOno08a} for reviews).
As
one of the applications of this structure behind the mock theta
functions,
we employ the method of
Bringmann and Ono~\cite{BrinKOno06a,BrinKOno08a}
(see also Ref.~\citenum{Bruin02Book}) in this paper to derive an 
exact
formula for the Fourier coefficients of
the elliptic genus for the K3 surface 
which counts the number of non-BPS (massive) representations. 
Analogous computations of  
Fourier coefficients of the partition function of three-dimensional
gravity   using the Poincar{\'e} series
were discussed in
Refs.~\citenum{DijMalMooVer00a,GMoo06a,MaloWitt07a,MansMoor07a}. 
In these papers Jacobi forms are considered instead of mock theta
functions.

We shall also  give an
exact  formula for the number of non-BPS
representations  for
the ALE space, which is a degenerate limit of the K3 surface.
The ALE spaces,
or the asymptotically locally Euclidean spaces,
are 
hyperK{\"a}hler 4-manifolds, and are constructed from
resolutions of
the Kleinian singularities.

We also would like to  point  out that the non-holomorphic partner of
the level-$1$ 
superconformal
characters  
is  connected to the
Witten--Reshetikhin--Turaev (WRT)
invariant  for  $3$-manifold
associated with the $D_4$-type  singularity.

This paper is organized as follows.
In section~\ref{sec:SCFT_mock} we recall a relationship
between the
$\mathcal{N}=4$
superconformal characters and the mock theta functions studied in
Ref.~\citenum{EguchiHikami08a}.
We briefly review the elliptic genus of the K3 surfaces in
section~\ref{sec:elliptic_genus}.
We show how to decompose  the elliptic genus in terms of the
superconformal characters.
In section~\ref{sec:Poincare},
we introduce the Poincar{\'e} series, whose holomorphic part has the
same
Fourier
coefficients  as 
the number of non-BPS representations.
Following Bringmann and Ono, 
we compute the Fourier expansion of the Poincar{\'e} series,
and give an exact asymptotic formula.
We  numerically compute the asymptotic expansion to confirm the
validity of our analytic expressions.
In section~\ref{sec:Chern-Simons} we recall a fact
that a limiting value of
the non-holomorphic part of the harmonic Maass form is related to the
SU($2$) WRT invariant for the Seifert manifold $M(2,2,2)$.
The last section is devoted to concluding remarks.
\section{
  The $\mathcal{N}=4$ Superconformal Algebras and Mock Theta
  Functions}
\label{sec:SCFT_mock}

The $\mathcal{N}=4$ superconformal algebra is generated by
the energy-momentum tensor, $4$ supercurrents, and a triplet of
currents which constitute the affine Lie algebra $SU(2)_k$.
The central charge is quantized to $c= 6 \, k$, and the unitary
highest weight state is labeled by the conformal weight $h$ and the
isospin $\ell$;
\begin{align*}
  L_0 \,
  \left| \Omega \right\rangle
  & =
  h \,
  \left| \Omega \right\rangle ,
  \\[2mm]
  T_0^3 \,
  \left| \Omega \right\rangle 
  & =
  \ell \, 
  \left| \Omega \right\rangle .
\end{align*}
where
$h \geq k/4$ and $0 \leq \ell \leq k/2$ in the Ramond sector.
The character of a representation is given by
\begin{equation}
  \ch_{k,h,\ell} ( z; \tau)
  =
  \Tr_{\mathcal{H}}
  \left(
    \E^{2  \, \pi \,  \I \,  z \,  T_0^3} \,
    q^{L_0 - \frac{c}{24}}
  \right) ,
\end{equation}
where $q=\E^{2 \, \pi \, \I \, \tau}$ with $\tau\in\mathbb{H}$, and
$\mathcal{H}$ denotes the Hilbert space of the representation.

There exist two types of representations in
the $\mathcal{N}=4$ superconformal
algebra~\cite{EgucTaor86a,EgucTaor88a}: massless (BPS) and massive
(non-BPS) representations. 
In the Ramond sector, their character formulas are given as follows;
\begin{itemize}
\item massless representations
  ($h=\frac{k}{4}$, and $\ell=0,\frac{1}{2}, \dots, \frac{k}{2}$),
  \begin{equation}
    \ch^R_{k,\frac{k}{4},\ell}(z;\tau)
    =
    \frac{\I}{\theta_{11}(2 z; \tau)} \cdot
    \frac{
      \left[ \theta_{10}(z;\tau) \right]^2
    }{
      \left[ \eta(\tau) \right]^3
    }
    \sum_{\varepsilon=\pm 1} \sum_{m \in \mathbb{Z}}
    \varepsilon \,
    \frac{
      \E^{4 \pi \I \varepsilon \left( (k+1)m+\ell \right) z}
    }{
      \left(
        1+ \E^{- 2 \pi \I \varepsilon z} \, q^{-m}
      \right)^2
    } \,
    q^{(k+1) m^2 + 2 \ell m} .
    \label{define_massless_ch}
  \end{equation}
  See Appendix for definitions of the Jacobi theta functions.
  
\item massive representations
  ($h > \frac{k}{4}$ and $\ell=\frac{1}{2}, 1, \dots, \frac{k}{2}$),
  \begin{equation}
    \label{massive_character}
    \ch^R_{k,h,\ell}(z;\tau)
    =
    q^{h - \frac{\ell^2}{k+1} - \frac{k}{4}} \,
    \frac{
      \left[ \theta_{10}(z;\tau) \right]^2
    }{
      \left[ \eta(\tau) \right]^3
    } \,
    \chi_{k-1, \ell - \frac{1}{2}}(z;\tau) ,
  \end{equation}
  where $\chi_{k,\ell}(z;\tau)$ denotes the affine SU($2$) character
  \begin{equation*}
    \chi_{k,\ell}(z;\tau)
    =
    \frac{
      \vartheta_{k+2,2\ell+1} - \vartheta_{k+2,-2\ell-1}
    }{
      \vartheta_{2,1} - \vartheta_{2,-1}
    }(z;\tau) ,
  \end{equation*}
  with the theta series defined by
  \begin{align*}
    \vartheta_{P,a}(z;\tau)
    & =
    \sum_{n \in \mathbb{Z}} q^{\frac{(2 P n +a)^2}{4 P}} \,
    \E^{2 \pi \I z (2 P n +a)} .
  \end{align*}

\end{itemize}

Characters in other sectors are related to each other by the spectral flow;
\begin{equation*}
  \begin{array}{ccc}
    R: z +\frac{\tau}{2} & \leftrightarrow & \widetilde{R}: z+\frac{1+\tau}{2}
    \\
    \updownarrow & & \updownarrow
    \\
    NS: z & \leftrightarrow & \widetilde{NS}: z+\frac{1}{2}
  \end{array}
\end{equation*}
Hereafter we study the theory at level $k=1$ 
($c=6$)
in the $\widetilde{R}$-sector,
where the
massless character
is  given by 
\begin{gather}
  \label{massless_ch_1}
  \ch_{k=1,h=\frac{1}{4},\ell=0}^{\widetilde{R}}(z;\tau)
  =
  \frac{\I}{\theta_{11}(2 z;\tau)} \cdot
  \frac{
    \left[ \theta_{11}(z;\tau) \right]^2
  }{
    \left[ \eta(\tau) \right]^3
  }
  \sum_{m \in \mathbb{Z}}
  q^{2 m^2} \, \E^{8 \pi \I m z} \,
  \frac{1 + \E^{2 \pi \I z} \, q^m}{
    1 - \E^{2 \pi \I z} \, q^m} .
\end{gather}
It is   known that this formula may be 
rewritten as~\cite{EgucTaor88b}
\begin{equation}
  \label{massless_ch_M}
  \ch_{k=1,h=\frac{1}{4},\ell=0}^{\widetilde{R}}(z;\tau)
  =
  \frac{
    \left[    \theta_{11}(z;\tau) \right]^2
  }{
    \left[ \eta(\tau) \right]^3
  } \,
  \mu (z;\tau) ,
\end{equation}
where $\mu(z;\tau)$ is a Lerch sum defined by
\begin{equation}
  \label{define_M}
  \mu(z;\tau)
  =
  \frac{
    \I \, \E^{\pi \I z} 
  }{
    \theta_{11}(z;\tau)}
  \sum_{n \in \mathbb{Z}}
  (-1)^n \,
  \frac{
    q^{\frac{1}{2} n(n+1)} \, \E^{2 \pi \I  n z}
  }{
    1 - q^n \, \E^{2 \pi \I z}
  } .
\end{equation}
Note that the massless characters fulfill an identity 
\begin{align}
  \label{recursion_ch_R_prime}
  \ch^{\widetilde{R}}_{k=1,h=\frac{1}{4},\ell=\frac{1}{2}}(z;\tau)
  +
  2 \,
  \ch^{\widetilde{R}}_{k=1,h=\frac{1}{4},\ell=0}(z;\tau)
  & =
  q^{-\frac{1}{8}} \,
  \frac{
    \left[ \theta_{11}(z;\tau) \right]^2
  }{
    \left[ \eta(\tau) \right]^3
  } 
  \\
  & =
  \lim_{h \searrow \frac{1}{4}}
  \ch^{\widetilde{R}}_{k=1,h,\ell=\frac{1}{2}}(z;\tau) ,
  \nonumber
\end{align}
which shows that the non-BPS representation decomposes 
into a sum of BPS representations at the unitarity bound.

As shown in Ref.~\citenum{Zweg02Thesis},
we can complete the Lerch sum $\mu(z;\tau)$ to a Jacobi-like form
$\widehat{\mu}(z;\tau)$ as
\begin{equation}
  \label{define_1_M_hat}
  \widehat{\mu}(z;\tau)
  =
  \mu(z;\tau)
  - \frac{1}{2} \, R(\tau) .
\end{equation}
Here $R(\tau)$ denotes a non-holomorphic function defined by
\begin{equation}
  \label{define_non-holomorphic}
  R(\tau)
  =
  \sum_{n \in \mathbb{Z}} (-1)^n \,
  \left[
    \sign\left(n+\frac{1}{2}\right)
    -
    E\left(
     \left( n + \frac{1}{2}
     \right) \,
     \sqrt{2 \, \Im \tau}
    \right)
  \right]
  \, q^{-\frac{1}{2} \left( n + \frac{1}{2} \right)^2} .
\end{equation}
$E(z)$ denotes the error function given by 
\begin{align}
  \label{define_error_function}
  E(z)
  = 2 \int_0^z \E^{- \pi u^2} \mathrm{d} u 
  =
  1 - \erfc \left(  \sqrt{\pi} \, z \right) .
\end{align}
The function $\widehat{\mu}(z;\tau)$ transforms like a Jacobi
form~\cite{EichZagi85}
as follows;
\begin{equation}
  \label{modular_M_hat}
  \begin{gathered}
    \widehat{\mu}(z;\tau)
    =
    -
    \sqrt{\frac{\I}{\tau}} \,
    \widehat{\mu}\left(
      \frac{z}{\tau} ; - \frac{1}{\tau}
    \right) ,
    \\[2mm]
    \widehat{\mu}(z;\tau+1) =
    \E^{-\frac{1}{4} \pi \I} \,
    \widehat{\mu}(z;\tau) ,
    \\[2mm]
    \widehat{\mu}(z+1;\tau)
    =
    \widehat{\mu}(z+\tau;\tau)
    =
    \widehat{\mu}(z;\tau)  .   
  \end{gathered}
\end{equation}

One can conclude from Ref.~\citenum{Zweg02Thesis}
that the non-holomorphic function~\eqref{define_non-holomorphic} is a
period integral of the third power of the Dedekind $\eta$-function,
\begin{equation}
  \label{R_and_eta}
  R(\tau)
  =
  \frac{1}{\sqrt{\I}}
  \int_{- \overline{\tau}}^{\I \infty}
  \frac{
    \left[ \eta(x) \right]^3
  }{
    \sqrt{
      x+\tau
    }} \,
  \mathrm{d} x .
\end{equation}
Then the function $\widehat{\mu}(z;\tau)$ satisfies
\begin{equation}
  \label{diff_mu_hat_eta}
  \frac{\partial}{\partial \overline{\tau}}
  \widehat{\mu}(z;\tau)
  =
  \frac{\I}{2} \,
  \frac{
    \left[ \eta(- \overline{\tau}) \right]^3
  }{
    \sqrt{
      2 \, \Im \tau
    }
  } ,
\end{equation}
and 
\begin{equation}
  \label{differential_Maass}
  \left( \Im \tau \right)^{\frac{3}{2}}
  \frac{\partial}{\partial \tau}  \sqrt{\Im \tau} 
  \frac{\partial}{\partial \overline{\tau}} \,
  \widehat{\mu}(z;\tau)
  = 0 .
\end{equation}
Here the derivatives in $\tau=u+\I \, v$ are defined by 
\begin{align*}
  \frac{
    \partial}{\partial
    \tau
  } & =
  \frac{1}{2} \,
  \left(
    \frac{\partial}{\partial u}
    -
    \I \,
    \frac{\partial}{\partial v}
  \right) ,
  &
  \frac{
    \partial}{\partial
    \overline{\tau}
  } & =
  \frac{1}{2} \,
  \left(
    \frac{\partial}{\partial u}
    +
    \I \,
    \frac{\partial}{\partial v}
  \right) .
\end{align*}
Thus the function $\widehat{\mu}(z;\tau)$ is a harmonic Maass form
with weight $1/2$ and has $\left[\eta(\tau)\right]^3$ as
its ``shadow'' 
according to the terminology of Zagier~\cite{Zagier08a}.
Similar $q$-series was studied in recent work on the Donaldson
invariant~\cite{MalmKOno08a}.
We note that the 
above differential equation~\eqref{differential_Maass}
reduces to
\begin{equation}
  \label{Laplacian_mu_hat}
  \Delta_{\frac{1}{2}} \, \widehat{\mu}(z;\tau) =0,
\end{equation}
where $\Delta_k$ is the hyperbolic Laplacian of weight $k$,
\begin{equation}
  \label{define_Laplacian}
  \Delta_k =
  -v^2 \, \left( \frac{\partial^2}{\partial u^2}
    +
    \frac{\partial^2}{\partial v^2}
  \right)
  + \I \, k \, v \,
  \left(
    \frac{\partial}{\partial u}
    +
    \I \,    \frac{\partial}{\partial v}
  \right)    .
\end{equation}


\section{Elliptic Genus of K3 Surface  and
  Harmonic Maass Form}
\label{sec:elliptic_genus}

\subsection{Decomposition of Elliptic Genus into $\mathcal{N}$=4 Characters}

The elliptic genus of the Calabi-Yau manifold $X$
with complex dimension $c/3$ is identified
as~\cite{EWitt87a}
\begin{equation}
  Z_X(z;\tau)
  =
  \Tr_{\mathcal{H}^R \otimes \mathcal{H}^R}
  (-1)^F \,
  \E^{2 \, \pi \, \I \, T_0^3 \, z} \,
  q^{L_0 - \frac{c}{24}} \,
  \overline{q}^{\overline{L}_0 - \frac{c}{24}} ,
\end{equation}
where $(-1)^F
=
\E^{\pi \, \I \, \left( T_0^3 -
    \overline{T}_0^3\right) }$,
and $\mathcal{H}^R$ is the Hilbert space on the Ramond sector.
In the case of K3 surface,
it is known that~\cite{EguOogTaoYan89a,KawaYamaYang94a}
\begin{equation}
  \label{genus_K3}
  Z_{K3}(z;\tau)
  =
  8 \,
  \left[
    \left(
      \frac{\theta_{10}(z;\tau)}{\theta_{10}(0;\tau)}
    \right)^2
    +
    \left(
      \frac{\theta_{00}(z;\tau)}{\theta_{00}(0;\tau)}
    \right)^2
    +
    \left(
      \frac{\theta_{01}(z;\tau)}{\theta_{01}(0;\tau)}
    \right)^2
  \right] .
\end{equation}
To rewrite the elliptic genus in terms of the 
level-$1$ 
characters 
we introduce the function $J(z;w;\tau)$   by
\begin{align}
  J(z; w; \tau)
  & =
   \frac{
     \left[ \theta_{11}(z;\tau) \right]^2
   }{
     \left[ \eta(\tau) \right]^3
   }
   \,
  \left(
    \widehat{\mu}(z;\tau) -
    \widehat{\mu}(w;\tau) 
  \right)
  \nonumber
  \\
  & =
  \frac{
    \left[ \theta_{11}(z;\tau) \right]^2
  }{
    \left[ \eta(\tau) \right]^3
  }
  \,
  \left(
    {\mu}(z;\tau) -
    {\mu}(w;\tau) 
  \right)
  \nonumber
  \\
  & = 
  \ch_{k=1,h=\frac{1}{4},\ell=0}^{\widetilde{R}}(z;\tau)
  -
  \frac{
    \left[ \theta_{11}(z;\tau) \right]^2
  }{
    \left[ \eta(\tau) \right]^3
  }  \,
  {\mu}(w;\tau)   .
  \label{define_lowest_T}
\end{align}
Note that
\begin{equation}
  \label{J_zero}
  J(z;z;\tau) =0 .
\end{equation}
It was shown~\cite{EguchiHikami08a} that $J(z;w;\tau)$ behaves like a 
2-variable Jacobi form~\cite{EichZagi85} under the modular
transformation;
\begin{equation}
  \begin{gathered}
    J(z;w;\tau)
    =
    \E^{-2 \pi \I \frac{z^2}{\tau}} \,
    J \left(
      \frac{z}{\tau} ; \frac{w}{\tau}; - \frac{1}{\tau}
    \right) ,
    \\[2mm]
    J(z+1; w; \tau)
    =
    J(z; w; \tau+1)
    = J(z; w+\tau; \tau)
    = J(z; w; \tau) ,
    \\[2mm]
    J(z+\tau; w; \tau)
    =
    q^{-1} \, \E^{- 4 \pi \I z} \,
    J(z;w;\tau) .
  \end{gathered}
\end{equation}
These modular properties together with~\eqref{J_zero}
show that
$J(z;w;\tau)$ with  $w$ at the half-periods 
$\left\{ \frac{1}{2},
  \frac{1+\tau}{2},
  \frac{\tau}{2}
\right\}$
is given by Jacobi theta functions as follows;
\begin{equation}
  \label{1_T_theta_2}
  \begin{aligned}
    J \left( z;\frac{1}{2}; \tau \right)
    & =
    \left(
      \frac{\theta_{10}(z;\tau)}{
        \theta_{10}(0;\tau)}
    \right)^2 , 
    \\
    J \left( z;\frac{1+\tau}{2}; \tau \right)
    & =
    \left(
      \frac{\theta_{00}(z;\tau)}{
        \theta_{00}(0;\tau)}
    \right)^2,
    \\
    J \left( z;\frac{\tau}{2}; \tau \right)
    &=
    \left(
      \frac{\theta_{01}(z;\tau)}{
        \theta_{01}(0;\tau)}
    \right)^2.
  \end{aligned}
\end{equation}
Note that the
Lerch sums introduced in Ref.~\citenum{EgucTaor88b} are proportional to 
$\mu(w;\tau)$ at 
$w=\frac{1}{2},\frac{1+\tau}{2}, \frac{\tau}{2}$,
respectively;
\begin{equation}
  \label{define_h_from_H}
  \begin{aligned}
    h_2(\tau)
    \equiv
    \frac{\mu \left( \frac{1}{2};\tau \right)}{
      \eta(\tau)}
    & =
    \frac{1}{
      \eta(\tau) \, \theta_{10}(0;\tau)}
    \sum_{n \in \mathbb{Z}}
    \frac{q^{\frac{1}{2} n(n+1)}}{
      1+ q^n} ,
    \\
    h_3(\tau)
    \equiv
    \frac{\mu  \left( \frac{1+\tau}{2}; \tau \right)}{
      \eta(\tau)}
    & =
    \frac{1}{
      \eta(\tau) \, \theta_{00}(0;\tau)}
    \sum_{n \in \mathbb{Z}}
    \frac{q^{\frac{1}{2} n^2 - \frac{1}{8}}}{
      1+ q^{n-\frac{1}{2}}} ,
    \\
    h_4(\tau)
    \equiv
    \frac{\mu\left(\frac{\tau}{2}; \tau\right)}{
      \eta(\tau)}
    & =
    \frac{1}{
      \eta(\tau) \, \theta_{01}(0;\tau)}
    \sum_{n \in \mathbb{Z}} (-1)^n
    \frac{
      q^{\frac{1}{2}n^2 - \frac{1}{8}}}{
      1 - q^{n-\frac{1}{2}}} .
  \end{aligned}
\end{equation}

We thus find that,
using~\eqref{1_T_theta_2},
the elliptic
genus~\eqref{genus_K3}  is written as
\begin{equation*}
  Z_{K3}(z;\tau)
  =
  24 \,
  \ch_{k=1,h=\frac{1}{4},\ell=0}^{\widetilde{R}}(z;\tau)
  -
  8 \,
  \left[
    \frac{\theta_{11}(z;\tau)}{\eta(\tau)}
  \right]^2 \,
  \sum_{a=2,3,4} h_a(\tau) .
\end{equation*}
For later convenience let us  define
\begin{align}
  \label{define_sum_mu}
  \Sigma(\tau) & \equiv
  8 \hskip-3mm  \sum_{
    w \in \left\{
      \frac{1}{2} , \frac{1+\tau}{2} , \frac{\tau}{2}
    \right\}
  }\hskip-3mm 
  \mu(w; \tau)
  \\
  & =
  8 \, \eta(\tau)
  \sum_{a=2,3,4}   h_a(\tau)
  \nonumber
  \\
  &=
  q^{-\frac{1}{8}}\left(
    2 - \sum_{n = 1}^\infty A_n \, q^n
  \right) .
  \nonumber
\end{align}
Here coefficients $A_n$ are positive integers, and some of them are
computed as follows;
\begin{equation}
  \label{l1_coefficient_massive}
  \begin{array}{c|rrrrrrrrrrr}
    n & 1 & 2 & 3 & 4 & 5 & 6 & 7 & 8 & 9 & 10 &\cdots\\
    \hline
    A_n& 90 & 462 & 1540 & 4554 & 11592 & 27830 &
    61686 & 131100 & 265650 & 521136 &\cdots
  \end{array}
\end{equation}
Using the massive character~\eqref{massive_character} ($k=1,\ell=1/2$)
and the identity~\eqref{recursion_ch_R_prime} we conclude that the
elliptic genus~\eqref{genus_K3} for K3 surface is decomposed into a
sum of superconformal characters as~\cite{EguOogTaoYan89a}
\begin{equation}
  \label{decompose_K3}
  Z_{K3}(z;\tau)
  =
  20 \, \ch_{1,\frac{1}{4},0}^{\widetilde{R}}(z;\tau)
  -
  2 \,  \ch_{1,\frac{1}{4},\frac{1}{2}}^{\widetilde{R}}(z;\tau)
  +
  \sum_{n = 1}^\infty A_n \,
  \ch_{1,n+\frac{1}{4},\frac{1}{2}}^{\widetilde{R}}(z;\tau) .
\end{equation}
Here the first two terms are isospin $\ell=0$ and $1/2$ massless
representations,
and the last term denotes an infinite sum of massive representations.

\subsection{Elliptic Genus of ALE space}
The isospin-$\frac{1}{2}$ term in~\eqref{decompose_K3} comes
from $\mu \left(\frac{1}{2};\tau \right)$
and corresponds to the identity representation in the NS sector. It  
describes the gravity multiplet in string compactification on $K3$
surface.
When we decompactify K3 into an ALE space,
we decouple gravity.  Thus we
may drop $\mu \left(\frac{1}{2};\tau \right)$ from the elliptic genus
and consider~\cite{EgucSugaTaor07a,EgucSugaTaor08a}
\begin{equation}
  \label{genus_K3_decompose}
  Z_{K3, \text{decompactified}}(z;\tau)
  =
  8 \,
  \left[
    \left(
      \frac{\theta_{00}(z;\tau)}{\theta_{00}(0;\tau)}
    \right)^2
    +
    \left(
      \frac{\theta_{01}(z;\tau)}{\theta_{01}(0;\tau)}
    \right)^2
  \right] .
\end{equation}
When we define
\begin{align}
  \Sigma^\circ(\tau)
  & \equiv
  8 \, \mu \left( \frac{1}{2} ; \tau \right)
  \nonumber
  \\
  & =
  q^{-\frac{1}{8}} \,
  \left(
    2 - \sum_{n = 1}^\infty A_n^{\circ} \, q^n
  \right),
  \label{decompact_A}
\end{align}
we have a character decomposition of the elliptic genus as
\begin{equation}
  Z_{K3,\text{decompactified}}(z;\tau)
  =
  16 \, \ch_{1,\frac{1}{4},0}^{\widetilde{R}}(z;\tau)
  +
  \sum_{n = 1}^\infty  \left(
    A_n - A_n^{\circ}
  \right) \,
  \ch_{1,n+\frac{1}{4},\frac{1}{2}}^{\widetilde{R}}(z;\tau) .
\end{equation}
Here coefficients $A_n^\circ$ are integers, and some of them are
computed from~\eqref{define_M} as follows;
\begin{equation}
  \label{l1_coefficient_decompose}
  \begin{array}{c|rrrrrrrrrrr}
    n & 1 & 2 & 3 & 4 & 5 & 6 & 7 & 8 & 9 & 10 &\cdots\\
    \hline
    A_n^\circ& -6 & 14 & -28 & 42 & -56 & 86 &
    -138 & 188 & -238 & 336 &\cdots
  \end{array}
\end{equation}

It is known that
the K3 surface may be decomposed into a sum of $16$ $A_1$
spaces~\cite{DPage78a},
and the elliptic genus of
$A_1$ space is proposed as~\cite{EgucSugaTaor07a};
\begin{align}
  Z_{A_1}(z;\tau)
  & =
  \frac{1}{2} \,
  \left[
    \left(
      \frac{\theta_{00}(z;\tau)}{\theta_{00}(0;\tau)}
    \right)^2
    +
    \left(
      \frac{\theta_{01}(z;\tau)}{\theta_{01}(0;\tau)}
    \right)^2
  \right] 
  \\
  & =
  \ch_{1,\frac{1}{4},0}^{\widetilde{R}}(z;\tau)
  +
  \frac{1}{16} \,
  \sum_{n = 1}^\infty  \left(
    A_n - A_n^{\circ}
  \right) \,
  \ch_{1,n+\frac{1}{4},\frac{1}{2}}^{\widetilde{R}}(z;\tau) .
  \nonumber
\end{align}
Note that $(A_n-A_n^{\circ})/16$ are all positive integers.

\section{Poincar{\'e}--Maass Series and Character Decomposition}
\label{sec:Poincare}
\subsection{Multiplier System for Harmonic Maass Forms}

As the completion of
$\Sigma(\tau)$ defined in~\eqref{define_sum_mu}, we define 
\begin{equation}
  \label{define_complete_mu}
   \widehat{\Sigma}(\tau)  =
   8 \hskip-5mm \sum_{
    w \in \left\{
      \frac{1}{2} , \frac{1+\tau}{2} , \frac{\tau}{2}
    \right\}
  }
  \widehat{\mu}(w; \tau) ,
\end{equation}
which transforms as
\begin{eqnarray}
  \label{mu_sum_transform_1}
   && \widehat{\Sigma}\left( - \frac{1}{\tau}
    \right)
    =
    - \sqrt{\frac{\tau}{\I}} \,
    \widehat{\Sigma}(\tau),
    \\[2mm]
   && \widehat{\Sigma}(\tau+1)
    =
    \E^{-\frac{1}{4} \pi \I} \,
    \widehat{\Sigma}(\tau) . 
     \label{mu_sum_transform_2} 
\end{eqnarray}
One finds in~\eqref{mu_sum_transform_1} and~\eqref{mu_sum_transform_2}
that
the multiplier system for $\Sigma$ is a complex conjugate to that
of $\left[ \eta(\tau) \right]^3$.
We recall a transformation formula of
the Dedekind $\eta$-function
(\emph{e.g.}~\citep[Chapter 9]{HRadema73});
\begin{equation}
  \label{transform_Dedekind_eta}
  \eta \left( \gamma(\tau) \right)
  =
   \I^{-\frac{1}{2}} \,
   \E^{\frac{a+d}{12 \, c} \,  \pi \, \I
   - s(d,c) \, \pi \,  \I} \,
   \sqrt{ c \, \tau + d} \,
   \eta(\tau)  .
\end{equation}
Here
$\gamma =
\begin{pmatrix}
  a & b
  \\
  c & d
\end{pmatrix}
\in SL(2;\mathbb{Z})$ with $c > 0$, and
$s(d,c)$ is the Dedekind sum defined by
\begin{equation*}
  s(d,c)
  =
  \sum_{k \mod c}
  \Biggl( \!\!  \Biggl( \frac{k}{c} \Biggr) \!\! \Biggr) \,
  \Biggl( \!\!  \Biggl( \frac{k\, d}{c} \Biggr) \!\! \Biggr) ,
\end{equation*}
where
\begin{equation*}
  ( \! ( x ) \! ) =
  \begin{cases}
    \displaystyle
    x - \lfloor x \rfloor - \frac{1}{2}  ,
    &
    \text{for $ x \in \mathbb{R} \setminus \mathbb{Z}$,}
    \\[2mm]
    0 , &
    \text{for $x \in \mathbb{Z}$.}
  \end{cases}
\end{equation*}
We thus conclude that
the completion~\eqref{define_complete_mu} satisfies
\begin{equation}
  \label{mu_sum_Gamma}
  \widehat{\Sigma}
  \left(
    \gamma(\tau) 
  \right)
  =
  \I^{\frac{3}{2}} \,
  \E^{- \frac{a+d}{4 c} \,  \pi \,  \I
    + 3   \, s(d, c) \, \pi \, \I} \,
  \sqrt{c \, \tau +d} \
  \widehat{\Sigma}(\tau) .
\end{equation}
\subsection{The Whittaker Function and
  the Poincar{\'e}--Maass Series}

The Whittaker functions~\cite{WhittWatso27},
$M_{\alpha,\beta}(z)$ and
$W_{\alpha,\beta}(z)$,
are solutions of the second order differential equation
\begin{equation}
  \left[
    \frac{\partial^2}{\partial  z^2}
    +
    \left(
      - \frac{1}{4} + \frac{\alpha}{z}
      +
      \frac{
        \frac{1}{4} - \beta^2
      }{z^2}
    \right)
  \right] \, w(z) = 0 .
\end{equation}
We have
\begin{equation*}
  W_{\alpha,\beta}(z)
  =
  \frac{
    \Gamma(-2 \, \beta)
  }{
    \Gamma \left( \frac{1}{2} - \alpha - \beta \right)
  } \,
  M_{\alpha,\beta}(z)
  +
  \frac{
    \Gamma(2 \, \beta)
  }{
    \Gamma \left( \frac{1}{2} - \alpha + \beta \right)
  } \,
  M_{\alpha, -\beta}(z) ,
\end{equation*}
and 
\begin{equation*}
  \begin{aligned}
    M_{\alpha,\beta}(z)
    & \underset{\Re z \to +\infty}{\sim}
    \frac{
      \Gamma(1+2 \, \beta)
    }{
      \Gamma
      \left(
        \frac{1}{2}
        - \alpha +        \beta 
      \right)
    } \,
    \E^{\frac{z}{2}} \,
    z^{- \alpha} ,
    \\[2mm]
    W_{\alpha,\beta}(z)
    &
    \underset{\Re z \to +\infty}{\sim}
    \E^{-\frac{z}{2}} \, z^\alpha .
  \end{aligned}
\end{equation*}
Following Ref.~\citenum{Bruin02Book}, we define functions
$\mathcal{M}^k_s(v)$ and
$\mathcal{W}^k_s(v)$
in terms of the
Whittaker functions as
\begin{equation}
  \begin{aligned}
    \mathcal{M}_s^k(v)
    & =
    \left| v \right|^{- \frac{k}{2}} \,
    M_{\frac{k}{2} \sign(v), s - \frac{1}{2}}
    \left( \left| v \right| \right) ,
    \\[2mm]
    \mathcal{W}^k_s(v)
    &=
    \left| v \right|^{-\frac{k}{2}} \,
    W_{\frac{k}{2} \, \sign(v), s - \frac{1}{2}}
    \left( \left| v \right| \right) .
  \end{aligned}
\end{equation}
Some of them are  given as follows;
\begin{align*}
  \mathcal{W}^{\frac{1}{2}}_{\frac{3}{4}}(v)
  & =
  \E^{- \frac{v}{2}},
  \\[2mm]
  \mathcal{W}^{\frac{1}{2}}_{\frac{3}{4}}(-v)
  & =
  \sqrt{\pi} \,
  \left(
    1 - E\left( \sqrt{\frac{v}{\pi}} \right)
  \right) \,
  \E^{ \frac{v}{2}},
  \\[2mm]
  \mathcal{M}^{\frac{1}{2}}_{\frac{3}{4}}(-v)
  & =
  \frac{\sqrt{\pi}}{2} \,
  E \left(\sqrt{\frac{v}{\pi}}\right) \,
  \E^{ \frac{v}{2}},
\end{align*}
where we assume $v>0$,
and the error function 
$E(z)$ is defined in~\eqref{define_error_function}.

We  define the function $\varphi_{-h,s}^k(\tau)$ for $h>0$ by
\begin{equation}
  \varphi_{-h,s}^k(\tau)
  =
  \mathcal{M}_s^k
  \left( -4 \, \pi \, h \, \Im (\tau) \right) \,
  \E^{- 2 \,  \pi \,  \I \,  h  \, \Re(\tau)} .
\end{equation}
Then  it becomes an eigenfunction of the
second order differential equation;
\begin{equation}
  \Delta_k \, \varphi_{-h,s}^k(\tau)
  =
  \left[
    s \, (1-s) + \frac{k}{2} \, \left( \frac{k}{2}-1 \right)
  \right] \,
  \varphi_{-h,s}^k(\tau) ,
\end{equation}
where $\Delta_k$ is the Laplacian defined in~\eqref{define_Laplacian}.
Note that at $\Im \tau\to +\infty$
\begin{equation*}
  \varphi^k_{-h,s}(\tau)
  \sim
  \frac{\Gamma (2 \, s)}{
    \Gamma
    \left(
      \frac{k}{2}+s
    \right)
  } \,
  q^{-h} .
\end{equation*}

By using  the function $\varphi_{-h,s}^k(\tau)$,
we introduce  the Poincar{\'e}--Maass series~\cite{Bruin02Book} as
\begin{equation}
  \label{define_Poincare-Maass}
  P_s(\tau)
  =
  \frac{2}{\sqrt{\pi}} \hskip-7mm 
  \sum_{ \gamma
    =
    {\scriptsize
      \begin{pmatrix}
        a & b
        \\
        c & d
      \end{pmatrix}
    }
    \in
    \Gamma_\infty \backslash \Gamma(1)}\hskip-7mm 
  \left[ \chi(\gamma) \right]^{-1} \,
  \frac{1}{\sqrt{c \, \tau+d}} \,
  \varphi_{-\frac{1}{8},s}^{\frac{1}{2}}\left( \gamma(\tau) \right) ,
\end{equation}
where the multiplier system is chosen 
to be
\begin{equation}
  \label{define_automorphy}
  \chi(\gamma)
  =
  \begin{cases}
    \I^{\frac{3}{2}} \,
    \E^{-\frac{a+d}{4 c}  \, \pi \,  \I+ 3 \, \pi \, \I \, s(d,c)},
    & \text{for $c > 0$,}
    \\[2mm]
    \E^{- \frac{b}{4} \, \pi \, \I} ,
    & \text{for $c=0$ and $d=1$.}
  \end{cases}
\end{equation}
We mean that
$\Gamma(1)=SL(2;\mathbb{Z})$, and 
$\Gamma_\infty$ is the stabilizer of $\infty$,
\begin{equation*}
  \Gamma_\infty =
  \left\{
    \begin{pmatrix}
      1 & n \\
      0 & 1
    \end{pmatrix}
    ~\Bigg|~
    n \in \mathbb{Z}
  \right\} .
\end{equation*}
We see that the Poincar{\'e}--Maass series
$P_{s}(\tau)$ transforms in the same way as 
$\widehat{\Sigma}(\tau)$~\eqref{mu_sum_transform_1},~\eqref{mu_sum_transform_2}
\begin{equation}
  P_{s}\left( \gamma(\tau) \right)
  =
  \chi(\gamma) \, \sqrt{c \, \tau+ d} \, P_{s}(\tau) ,
\end{equation}
and that, due to a commutativity of the Laplacian
$\Delta_{k}$~\eqref{define_Laplacian}  and the
$\gamma$ action,
it is an eigenfunction of the Laplacian
\begin{equation}
  \Delta_{\frac{1}{2}} P_s(\tau)
  =
  \left[
    s \, (1-s) - \frac{3}{16}
  \right] \, P_s(\tau) .
\end{equation}
At $\Im \tau \to +\infty$,  the asymptotic behavior of the
$M$-Whittaker function shows
that  the Poincar{\'e}--Maass series behaves exponentially as
$\E^{\pi \, \Im(\tau)/4}$.

Note that $P_s(\tau)$ is annihilated by the Laplacian
$\Delta_{\frac{1}{2}}$ like
$\widehat{\mu}(z;\tau)$~\eqref{Laplacian_mu_hat} when we set
$s=\frac{3}{4}$;
\begin{equation}
  \Delta_{\frac{1}{2}} \, P_{\frac{3}{4}}(\tau) = 0.
\end{equation}
These facts show that the Poincar{\'e}--Maass series
$P_{\frac{3}{4}}(\tau)$ is the harmonic Maass form with weight $1/2$.

We shall  compute
the Fourier coefficients 
of $P_s(\tau)$
in the form of Rademacher expansion.
By using the standard method,
we have
\begin{multline*}
  P_s(\tau)
  =
  \frac{4}{\sqrt{\pi}} \, \varphi^{\frac{1}{2}}_{-\frac{1}{8},s}(\tau)
  \\
  +
  \frac{4}{\sqrt{\pi}} \hskip-4mm
  \sum_{
    \substack{
      c\neq 0 \\
      \gamma
      \in \Gamma_\infty \backslash \Gamma(1) / \Gamma_\infty
    }}\hskip-4mm 
  \sum_{n \in \mathbb{Z}}
  \left[
    \chi
    \left( \gamma \,
      \begin{pmatrix}
        1 & n \\
        0 & 1
      \end{pmatrix}
    \right)
  \right]^{-1}
  \frac{1}{\sqrt{
      c \, \left(\tau + n \right) + d
    }}
  \,
  \varphi^{\frac{1}{2}}_{- \frac{1}{8},s}
  \left(
    \gamma \,
    \begin{pmatrix}
      1 & n \\
      0 & 1
    \end{pmatrix}
    (\tau)
  \right) .
\end{multline*}
Using
$\gamma(\tau)
=\frac{a}{c} - \frac{1}{c \, (c \, \tau +d )}$,
the second term
up to an overall factor
is rewritten as
\begin{multline*}
  \sum_{\substack{
      c>0 \\
      \gamma \in \Gamma_\infty
      \backslash
      \Gamma(1)  
      /
      \Gamma_\infty
    }}
  \frac{1}{\sqrt{c}} \,
  \left[ \chi(\gamma) \right]^{-1} \,
  \E^{-
    \frac{a}{4 \,   c} \, \pi \, \I
  }
  \\
  \times
  \sum_{n \in \mathbb{Z}}
  \left[
    \chi \left(
      \begin{pmatrix}
        1 & n \\
        0 & 1
      \end{pmatrix}
    \right)
  \right]^{-1} \,
  \frac{1}{
    \sqrt{\tau + n + \frac{d}{c}}
  } \,
  \mathcal{M}_s^{\frac{1}{2}}
  \left(
    - {\pi\over 2} \, \frac{\Im (\tau)}{
      c^2 \,
      \left|
        \tau + n+\frac{d}{c}
      \right|^2
    }
  \right) \,
  \E^{ \frac{1}{4 \, c^2} \,
    \Re
    \left(
      \frac{1}{\tau +n+\frac{d}{c}}
    \right) \,
    \pi \I
  } .
\end{multline*}
We then  apply  a Fourier transformation
formula~\cite{Hejhal83Book,Bruin02Book}
\begin{multline}
  \sum_{n \in \mathbb{Z}}
  \frac{1}{
    \sqrt{\tau +n}} \,
  \mathcal{M}^{\frac{1}{2}}_s \left(
    -4 \, \pi \, h \, 
    \frac{\Im (\tau) }{
      c^2 \, \left| \tau + n \right|^2}
  \right) \,
  \E^{2 \, \pi \,  \I \,  h^\prime \, n
    +2 \,  \pi \,  \I \,  \frac{h}{c^2}
    \Re \left( \frac{1}{\tau+n} \right)
  }
  \\
  =\sum_{n \in \mathbb{Z}} \,
  a_n\left(\Im (\tau) \right) \,
  \E^{2 \, \pi \,  \I \, 
    \left( n-h^\prime \right)
    \, \Re(\tau)
  } ,
\end{multline}
where the Fourier coefficients $a_n(v)$  are given as follows;
\begin{itemize}
\item for $n>h^\prime$,
  \begin{multline*}
    a_n(v)
    =
    \frac{1}{\sqrt{\I}} \,
    \left( \frac{h}{4 \, \pi \, c^2 \, v} \right)^{\frac{1}{4}}
    \,
    \frac{
      \Gamma(2 \, s)}{
      \Gamma \left( s+\frac{1}{4} \right)
    }
    \\
    \times
    \frac{2 \, \pi 
    }{
      \sqrt{n-h^\prime}} \,
    W_{\frac{1}{4}, s - \frac{1}{2}}
    \left(
      4 \, \pi \,
      \left( n - h^\prime \right) \, v 
    \right) \,
    I_{2s-1}
    \left(
      \frac{4 \, \pi}{
        \left| c \right|} \,
      \sqrt{\left( n-h^\prime \right) \, h}
    \right) ,
  \end{multline*}

\item for $n=h^\prime$,
  \begin{equation*}
    a_n(v)
    =
    \frac{1}{\sqrt{\I}} \,
    \frac{
      2^{\frac{3}{2}} \, \pi^{s+\frac{3}{4}} \,
      \Gamma(2 \, s)
    }{
      (2 \, s -1) \,
      \Gamma \left(s+\frac{1}{4} \right) \,
      \Gamma \left(s-\frac{1}{4} \right)
    } \,
    \frac{h^{s-\frac{1}{4}}}{
      \left| c \right|^{2s - \frac{1}{2}} \,
      v^{s-\frac{3}{4}}
    } ,
  \end{equation*}

\item for $n<h^\prime$,
  \begin{multline*}
    a_n(v)
    =
    \frac{1}{\sqrt{\I}} \,
    \left( \frac{h}{4 \, \pi \, c^2 \, v} \right)^{\frac{1}{4}}
    \,
    \frac{
      \Gamma(2 \, s)}{
      \Gamma \left( s - \frac{1}{4} \right)
    }
    \\
    \times
    \frac{2 \, \pi
    }{
      \sqrt{h^\prime - n}} \,
    W_{- \frac{1}{4}, s - \frac{1}{2}}
    \left(
      4 \, \pi \,
      \left(  h^\prime - n \right) \, v 
    \right) \,
    J_{2s-1}
    \left(
      \frac{4 \, \pi}{
        \left| c \right|} \,
      \sqrt{\left( h^\prime - n \right) \, h}
    \right) .
  \end{multline*}
\end{itemize}
Here 
the (modified) Bessel function,  $I_\alpha(z)$ and $J_\alpha(z)$,
satisfy
\begin{gather*}
  \begin{aligned}
    I_\alpha(z)
    &\underset{z\to 0}{\sim} \frac{1}{
      \Gamma(\alpha+1)}
    \left( \frac{z}{2} \right)^{\alpha} \,
     ,
    &
    I_\alpha(z)
    & \underset{
      \left|z \right| \to \infty
    }{\sim}
    \frac{1}{
      \sqrt{2 \, \pi \, z}} \,
    \E^z ,
    \\
    J_\alpha(z)
    &\underset{z\to 0}{\sim}\frac{1}{
      \Gamma(\alpha+1)}
    \left( \frac{z}{2} \right)^{\alpha} \,
     ,
    & 
    J_\alpha(z)
    & \underset{
      \left|z \right| \to \infty
    }{\sim}
    \sqrt{\frac{2}{\pi \, z}} \,
    \cos
    \left(
      z- \frac{\alpha}{2} \, \pi - \frac{1}{4} \, \pi
    \right) .
  \end{aligned}
\end{gather*}
As a result, we obtain the Fourier expansion of the
Poincar{\'e}--Maass series as follows;
\begin{multline}
  \label{Fourier_Poincare-series}
  P_{\frac{3}{4}}(\tau)
  =
  2 \, E\left(\sqrt{\frac{\Im(\tau)}{2}}\right) \, q^{-\frac{1}{8}}
  \\
  -
  \sum_{\substack{
      n \in \mathbb{Z}
      \\
      n \ge 1
    }}
  \frac{4 \, \pi}{
    \left( 8 \, n -1 \right)^{\frac{1}{4}}
  } \,
  \left[
    \sum_{c > 0} \frac{1}{c} \,
    I_{\frac{1}{2}}
    \left( \frac{\pi}{2 \, c} \,  \sqrt{8 \, n -1} \right)
    \sum_{\substack{
        d \mod c
        \\
        (c,d)=1
      }}
    \E^{-3 \, \pi \, \I \, s(d,c)
      + 2 \,  \pi \,  \I \,  n \,  \frac{d}{c}}
  \right]\, q^{n-\frac{1}{8}}
  \\
  -
  \sum_{\substack{
      n \in \mathbb{Z}
      \\
      n \le 0
    }}
  \frac{4 \, \pi}{
    \left(
      1 - 8 \, n
    \right)^{\frac{1}{4}}
  } \,
  \left[
    1 - E \left( \sqrt{\frac{(1- 8 \, n) \, \Im (\tau)}{2} } \right)
  \right]
  \\
  \times
  \left[
    \sum_{c>0}
    \frac{1}{c} \, J_{\frac{1}{2}}
    \left(
      \frac{\pi}{2 \, c} \,   \sqrt{1-8 \, n}
    \right) \,
    \sum_{\substack{
        d \mod c
        \\
        (c,d)=1
      }}
    \E^{-3 \, \pi \, \I \,  s(d,c)
      + 2  \,  \pi \, \I \, \frac{d}{c} \, n 
    }
  \right] \,
  q^{n - \frac{1}{8} } .
\end{multline}
Proving the convergence of the above series in $c>0$ is a somewhat
difficult issue.
Such Poincar{\'e}--Maass series appeared in the
Andrews--Dragonette identity,
and  its convergence  was proved by Bringmann
and Ono~\citep[Section 4]{BrinKOno06a} by use of properties of the
Kloosterman sums and Sali{\'e} sums.
In our case,
we recall  that the sum involved
in~\eqref{Fourier_Poincare-series} can be rewritten as
\begin{equation}
  \sum_{\substack{
      d \mod c
      \\
      (c,d) =1
    }}
  \E^{-3 \, \pi \, \I \,  s(d,c)
    + 2  \,  \pi \, \I \, \frac{d}{c} \, n 
  }
  =
  - \frac{\I \, \sqrt{c}}{2}
  \sum_{\substack{
      k \mod 4 c
      \\
      k^2 = -8 \, n+1 \mod 8 c
    }}
  \left(
    \frac{-4}{k}
  \right) \,
  \E^{\pi \,  \I \,  \frac{k}{2 c}} .
\end{equation}
The sum on the right hand side can be parameterized by use of binary
quadratic forms of discriminant $-8 \, n+1$,
and the method of Bringmann and Ono shows how this description
naturally forces sufficient cancellation to justify convergence.
In section~\ref{sec:numerical}
we present evidence for the
convergence of the series by 
numerically computing the truncated series at finite values of $c$.

Our claim is that
the Poincar{\'e}--Maass series~\eqref{define_Poincare-Maass}
coincides with
the completion~\eqref{define_complete_mu};
\begin{equation}
  \label{Poincare_and_Maass}
  \widehat{\Sigma}(\tau)
  = P_{\frac{3}{4}} (\tau) .
\end{equation}
Proof of this statement is analogous to that of the
Andrews--Dragonette identity by Bringmann and Ono~\cite{BrinKOno06a}.
We first look at the non-holomorphic part of
the Poincar{\'e}--Maass
series $P_{\frac{3}{4}}(\tau)$.
From the above expression~\eqref{Fourier_Poincare-series},  we obtain 
\begin{multline}
  \label{differential_Poincare}
  \I  \sqrt{ \Im \tau} \,
  \overline{
    \frac{
      \partial}{
      \partial \overline{\tau}}
    P_{\frac{3}{4}}(8 \, \tau)
  }
  \\
  =
  \sum_{n=0}^\infty
  \left[
    2 \, \delta_{n,0}+
    \left( 8 \, n + 1 \right)^{\frac{1}{4}} \,
    \sum_{c>0}
    \frac{ 4 \, \pi }{c} \, J_{\frac{1}{2}}
    \left(
      \frac{\pi}{2 \, c} \,   \sqrt{8 \, n + 1}
    \right) \,
    \sum_{\substack{
        d \mod c
        \\
        (c,d)=1
      }}
    \E^{3 \, \pi \, \I \,  s(d,c)
      + 2  \,  \pi \, \I \, \frac{d}{c} \, n 
    }
  \right] \,
  {q}^{8 \, n+1} .
\end{multline}
As was proved by
Bruinier and Funke~\citep[Proposition 3.2]{BruinFunke04a},
the left hand side 
is a
weight-$3/2$ cusp form on $\Gamma_0(64)$ with a trivial character.
On the other hand
we have from~\eqref{diff_mu_hat_eta} that
\begin{align}
  \label{differential_eta_triple}
  \I \,
  \sqrt{
    \Im \tau
  } \,
  \overline{
    \frac{\partial}{\partial \overline{\tau}}
    \widehat{\Sigma}(8 \, \tau)
  }
  & =
  24 \, 
  \left[ \eta
    \left(  8 \, {\tau} \right)
  \right]^3 
  \\
  & =
  24 \,
  \left[
    {q} 
    - 3 \, {q}^9
    + 5 \, {q}^{25}
    - 7 \, {q}^{49}
    + 9 \, {q}^{81}
    - 11 \,{q}^{121}
    + 13 \,{q}^{169}
    - \cdots
  \right] ,
  \nonumber
\end{align}
which is also
a weight-$3/2$ cusp  form  
on $\Gamma_0(64)$ with a trivial character.
According to the dimension formulas for spaces of half-integral weight
modular forms~\cite{CoheOest77a} (see also Ref.~\citenum{KOno04Book}),
the dimension of cusp form on $\Gamma_0(64)$ is 1.
We thus see that
$\widehat{\Sigma}(8 \, \tau)$ 
is proportional to
$P_{\frac{3}{4}}(8 \, \tau)$.
Next by comparing the coefficients of $q^{-1}$,
we see that the holomorphic part of $P_{\frac{3}{4}}(8 \, \tau)$
coincides with that of $\Sigma(8 \, \tau)$ 
and hence we have the equality~\eqref{Poincare_and_Maass}.

Finally, we obtain an exact asymptotic expansion for
$A_n$~\eqref{l1_coefficient_massive} as
\begin{equation}
  \label{exact_A_expansion}
  A_n =
  \frac{4 \, \pi}{\left( 8 \, n -1 \right)^{\frac{1}{4}}} \,
  \sum_{c=1}^\infty
  \frac{
    1}{c} \,
  I_{\frac{1}{2}}\left( \frac{\pi \, \sqrt{8\, n -1}}{2 \, c} \right)
  \,
  \sum_{\substack{
      d \mod c \\
      (c,d)=1
    }}
  \E^{-3  \, \pi \, \I  \,  s(d,c) 
    + 2 \, \pi \, \I \,  \frac{d}{c}  \, n 
  } .
\end{equation}
The dominating contribution comes from the term $c=1$  in the above
expression,
\begin{equation}
  \label{asymptotics_A}
  A_n \sim
  \frac{4 \, \pi}{
    \left( 8 \, n - 1 \right)^{\frac{1}{4}}
  } \,
  I_{\frac{1}{2}}
  \left(
    \frac{\pi \, \sqrt{8 \, n -1}}{2}
  \right) .
\end{equation}
Substituting an explicit form of the Bessel function,
\begin{equation*}
  I_{\frac{1}{2}}(x)
  =
  \sqrt{\frac{2}{\pi \, x}} \,
  \sinh(x) ,
\end{equation*}
we have the Cardy type formula
\begin{equation}
\label{K3entropy}
  \log A_n
  \sim 2 \, \pi \,
  \sqrt{\frac{
      1}{2} \, \left( n - \frac{1}{8} \right)
  } .    
\end{equation}

\subsection{Non-Compact Case}
We shall next determine the Fourier coefficients
$A_n^\circ$~\eqref{decompact_A} which are related to the number of
non-BPS representations in the decompactified K3 surface.
We set the completion of $\Sigma^\circ(\tau)$ to be
\begin{equation}
  \widehat{\Sigma}^\circ(\tau)
  =
  8 \, \widehat{\mu}
  \left( \frac{1}{2} ; \tau \right) .
\end{equation}
Its  modular transformation formulae can be deduced
from~\eqref{modular_M_hat}.
Recalling the transformation formulae for the Jacobi theta function
$\theta_{10}(z;\tau)$
(\emph{e.g.}~\citep[Chapter 10]{HRadema73}),  we see that
for $\gamma \in \Gamma_0(2)$ with $c>0$
\begin{equation}
  \widehat{\Sigma}^\circ
  \left(
    \gamma(\tau) 
  \right)
  =
  \I^{\frac{3}{2}} \,
  \E^{- \frac{a+d}{4 c} \,  \pi \,  \I
  + 3 \, \pi \, \I \, s(d,c)} \,
  \sqrt{c \, \tau +d} \
  \widehat{\Sigma}^\circ(\tau) .
\end{equation}
Using the same argument with the above, we 
conclude that
it coincides with the
Poincar{\'e}--Maass series for
$\widehat{\Sigma}^\circ(\tau)$,
\begin{equation}
  \widehat{\Sigma}^\circ(\tau)
  =
  \frac{2}{\sqrt{\pi}} \,
  \sum_{\gamma \in
    \Gamma_\infty \backslash \Gamma_0(2)}
  \left[ \chi(\gamma) \right]^{-1} \,
  \frac{1}{\sqrt{c \, \tau+d}} \,
  \varphi_{-\frac{1}{8},\frac{3}{4}}^{\frac{1}{2}}\left( \gamma(\tau) \right) ,
\end{equation}
where the multiplier system
$\chi(\gamma)$ is given in~\eqref{define_automorphy}.
Correspondingly the Fourier coefficients
of $\widehat{\Sigma}^\circ(\tau)$ can be computed from the
Poincar{\'e}--Maass series, and
we obtain
\begin{equation}
  \label{exact_AC_expansion}
  A^\circ_n =
  \frac{4 \, \pi}{\left( 8 \, n -1 \right)^{\frac{1}{4}}} \,
  \sum_{\substack{
      c=1
      \\
      2 | c}
  }^\infty
  \frac{
    1}{c} \,
  I_{\frac{1}{2}}\left( \frac{\pi \, \sqrt{8\, n -1}}{2 \, c} \right)
  \,
  \sum_{\substack{
      d \mod c \\
      (c,d)=1
    }}
  \E^{-3  \, \pi \, \I \, s(d,c)
    + 2  \, \pi \, \I \,  \frac{d}{c}  \, n
  } .
\end{equation}
Asymptotic behavior of coefficients is given as
\begin{equation}
  \label{asymptotics_AC}
  A_n^\circ
  \sim
  (-1)^n \, \frac{2 \, \pi}{
    \left( 8 \, n - 1 \right)^{\frac{1}{4}}
  } \,
  I_{\frac{1}{2}}
  \left(
    \frac{\pi \,
      \sqrt{8 \, n -1}
    }{4}
  \right) .
\end{equation}
\subsection{Numerical Checks}
\label{sec:numerical}
\begin{figure}
  \centering
  \includegraphics[scale=1.0]{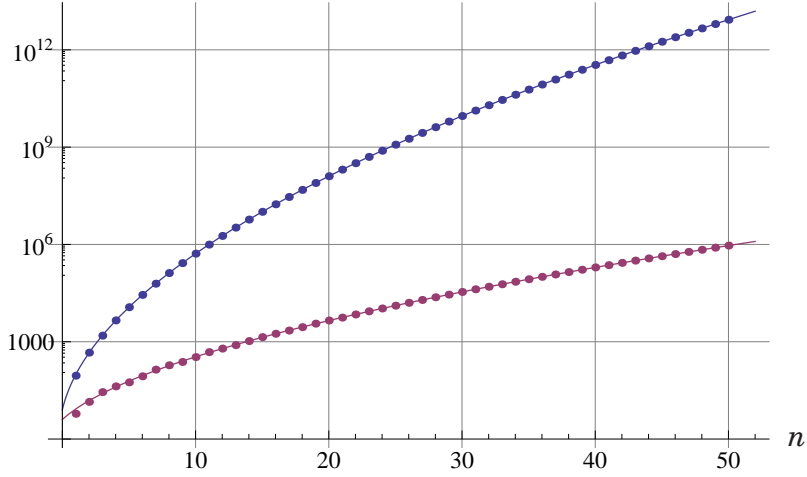}
  \caption{Exact values of
    $A_n$~\eqref{l1_coefficient_massive}
    and values of the asymptotic
    formula~\eqref{asymptotics_A} are
    given by blue dots and blue line, respectively.
    We have also plotted exact absolute values
    $|A_n^\circ|$~\eqref{l1_coefficient_decompose} together 
    with their asymptotic values~\eqref{asymptotics_AC} in red.
  }
  \label{fig:An}
\end{figure}

Now 
we present some results of numerical calculations and their comparison
with exact results in order to confirm the convergence of the
series~\eqref{Fourier_Poincare-series} 
and asymptotic formulas.
We have plotted the exact values of Fourier coefficients $A_n$ together 
with the values of the asymptotic 
formula~\eqref{asymptotics_A} in Fig.~\ref{fig:An}.
We have also presented the exact values of $A_n^\circ$
and compared  with the predictions given by 
modified  Bessel function~\eqref{asymptotics_AC}.
We find very good agreements.

In the tables we present more detailed results.
We have numerically computed
the  exact asymptotic series~\eqref{exact_A_expansion} of $A_n$ 
by truncating the infinite sum $\sum_{c=1}^\infty$
by a  finite number of terms.
For comparison we also list 
the exact  values  of $A_n$;
\begin{equation*}
  \begin{array}{r||r|rrr}
    n & \mbox{exact} & \mbox{leading} ;\,\eqref{asymptotics_A} &
    \text{sum of $5$ terms} 
    &
    \text{$20$ terms} 
    \\
    \hline \hline
    2 &
    462
    &
    453.018
    &
    462.026
    &
    462.427
    \\
    5 &
    11592
    &
    11662.495
    &
    11594.141
    &
    11592.421
    \\
    20 &
    126894174
    &
    126889894.140
    &
    126894174.078
    &
    126894173.718
    \\
    30 &
    9104078592
    &
    9104043456.138
    &
    9104078600.515
    &
    9104078592.403
    \\
    40 &
    342322413552
    &
    342322217629.135
    &
    342322413549.736
    &
    342322413551.574
    \\
    45 &
    1778826191324
    &
    1778826619936.736
    &
    1778826191295.658
    &
    1778826191322.367
  \end{array}
\end{equation*}
For the case of $A_n^\circ$,
we also present their exact values 
and values given by the truncation of the asymptotic 
expansion~\eqref{exact_AC_expansion} 
\begin{equation*}
  \begin{array}{r||r|rrr}
    n & \mbox{exact}& \mbox{leading} ;\,\eqref{asymptotics_AC}& \text{sum of $5$ terms}&
    \text{$10$ terms}
    \\
    \hline \hline
    5 &
    -56
    &
    -61.111
    &
    -56.544
    &
    -56.336
    \\
    20 &
    4510
    &
    4486.206
    &
    4511.303
    &
    4509.981
    \\
    21 &
    -5544
    &
    -5598.785
    &
    -5543.374
    &
    -5543.584
    \\
    40 &
    195888
    &
    195787.459
    &
    195888.432
    &
    195887.820
    \\
    60 &
    3772468
    &
    3772123.173
    &
    3772465.128
    &
    3772468.117
    \\
    100 &
    438370422
    &
    438366833.884
    &
    438370424.848
    &
    438370421.862
  \end{array}
\end{equation*}
One sees that
numerical results support the convergence of the
series~\eqref{Fourier_Poincare-series} 
the validity of our asymptotic 
formulae~\eqref{exact_A_expansion} and~\eqref{exact_AC_expansion}.

For its direct verification, we have
numerically  computed~\eqref{differential_Poincare} 
to obtain
\begin{multline*}
  \I  \sqrt{ \Im \tau} \,
  \overline{
    \frac{
      \partial}{
      \partial \overline{\tau}}
    P_{\frac{3}{4}}(8 \, \tau)
  }
  =
  23.851 \, {q}
  -72.0946 \,{q}^9
  +0.320386 \,{q}^{17}
  +
  119.083 \, {q}^{25}
  \\
  +
  0.295543 \, {q}^{33}
  +0.152477 \, {q}^{41}
  - 166.728 \, {q}^{49}
  -0.587912 \, {q}^{57}
  \\
  -0.0773375 \, {q}^{65}
  -0.652751 \, {q}^{73}
  + 213.397 \, {q}^{81}
  - 0.217745 \,{q}^{89}
  +\cdots .
\end{multline*}
Here
we have truncated the sum over $c$ after the first $800$ terms.
Though the convergence is slower than before, 
this agrees with~\eqref{differential_eta_triple}.

\section{Chern--Simons Theory}
\label{sec:Chern-Simons}

A key
in our analysis
is 
to complete the massless superconformal character $\mu(z;\tau)$
by adding
the non-holomorphic partner
$R(\tau)$~\eqref{define_non-holomorphic} so that the sum has a 
nice modular property.
Here we would like to
point out that $R(\tau)$ 
has its own meaning as a
topological quantum   invariant related to a simple
singularity.

The Witten invariant of 3-manifold $M$~\cite{EWitt89a} is
defined by the Chern--Simons path integral
\begin{equation}
  Z_k(M)
  =
  \int
  \E^{2 \, \pi \, \I \, k \, \CS(A)}
  \mathcal{D}A ,
\end{equation}
where the coupling constant
$k\in \mathbb{Z}$ denotes the level,
and
$A$ is a $G$-gauge
connection on the trivial bundle over $M$.
The Chern--Simons action $\CS(A)$ with gauge group $G$ is
\begin{equation*}
  \CS(A)
  =
  \frac{1}{8 \, \pi^2}
  \int\limits_{M}
  \Tr
  \left(
    A \wedge \mathrm{d} A
    + \frac{2}{3} \, A \wedge A \wedge A
  \right) .
\end{equation*}
See Ref.~\citenum{ResheTurae91a} for mathematically rigorous
definition of this quantum invariant
(Witten--Reshetikhin--Turaev invariant)
in terms of the quantum  invariants of links to be surgered.
When $M$ is the Poincar{\'e} homology sphere, it was shown that the 
Witten invariant
$Z_k(M)$  with gauge group $SU(2)$
can be regarded as a limiting value of the Eichler integral of 
a weight-$3/2$ modular form~\cite{LawrZagi99a}.
This correspondence  has been checked for other Seifert
manifolds~\cite{KHikami04b,KHikami05a}.

Let $M$ be the  Seifert manifold $M(2,2,2)$~\cite{JMiln75a}, which is a
spherical neighborhood around the isolated singularity of type-$D_4$;
\begin{equation*}
  M=
  \left\{ x^2 \, y + y^3 + z^2 = 0 \right\}
  \cap
  S^5 ,
\end{equation*}
where $S^5$ denotes a sufficiently small $5$-sphere around the origin.
It was shown~\cite{KHikami05a}
that the  Witten invariant  $Z_k(M)$ with SU($2$) gauge group is given by
(see also Refs.~\citenum{KHikami05c,KHikami05b})
\begin{equation}
  Z_k(M)
  =
  \frac{1}{2 \, \I} \,
  \sqrt{
    \frac{2}{k+2}
  } \,
  \E^{- \frac{1}{k+2} \, \pi \, \I} \,
  \left[
    1 - 
    \E^{- \frac{1}{4(k+2)} \pi \I} \,
    R
    \left(
      - \frac{1}{k+2}
    \right)
  \right] .
\end{equation}
Here $R(\tau)$ is the period integral of 
the third power of
the Dedekind $\eta$-function~\eqref{R_and_eta}, and we have a limiting
value in
$\tau\to-\frac{1}{k+2}$.
It was shown that the topological invariants, such as the Chern--Simons
invariant, the Reidemeister torsion, and the Ohtsuki invariant, of $M$
are given by the asymptotic expansion of $Z_k(M)$ in $k \to \infty$.

It should also be noted that
the non-holomorphic partner  of the
massless character $R(\tau)$ is related to the 
knot invariant for the torus link $T_{2,4}$~\cite{KHikami03a}
\begin{equation}
  \left\langle T_{2,4} \right\rangle_N
  =
  N \,
  \E^{- \frac{1}{4 N} \, \pi \, \I} \,
  R \left(
    - \frac{1}{N}
  \right) .
\end{equation}
Here
$\left\langle L \right\rangle_N$ denotes
a specific value of the $N$-colored Jones polynomial
for link $L$  at $q=\E^{2 \pi \I/N}$: this quantity receives renewed interests
from the viewpoint of  the volume conjecture 
raised by Kashaev~\cite{Kasha96b}, and
H.~Murakami and J.~Murakami~\cite{MuraMura99a}.

\section{Concluding Remarks}

As an application of our previous result~\cite{EguchiHikami08a} that
the coefficients of massive characters of the elliptic genera are the
holomorphic part of  harmonic Maass form,
we have obtained their Rademacher-type expansion by
computing the Fourier coefficients of the Poincar{\'e}--Maass series.

We note that in the elliptic genus the right-moving sector is fixed to
Ramond ground state  and thus the non-BPS states in the left-moving
sector are actually the overall half-BPS states  (BPS (non-BPS) in the
right (left) moving sector).
It is known that asymptotic increase of
the number of half-BPS states 
is related to the entropy of supersymmetric systems.
In fact the multiplicity factor $A_n$ behaves like an
exponential~\eqref{K3entropy} and we may identity 
\begin{equation}
  S=2 \, \pi \, \sqrt{{1\over 2} \, n}
\end{equation}
as the entropy of K3 surface.


Our  methods are applicable to higher level superconformal
algebras and higher-dimensional hyperK{\"a}hler manifolds. 
In general hyperK{\"a}hler manifolds with complex $2k$-dimensions we
find an entropy ~\cite{EguchiHikami09b}
\begin{equation}
  S=2 \, \pi \,  \sqrt{{k^2\over k+1} \, n}.
\end{equation}
If we consider the case of symmetric product of K3 surfaces
$K3^{[k]}$, the above entropy reproduces the black hole entropy of
string theory compactified on K3 surface at large $k$.

\section*{Acknowledgments}
One of the authors (KH) would like to thank 
M.~Kaneko
and in particular K.~Ono for 
his useful
communications on the issue of convergence of Poincar{\'e} series.
This work is supported in part by Grant-in-Aid from the Ministry of
Education, Culture, Sports, Science and Technology of Japan.

\appendix
\section{Jacobi Theta Functions}
The Jacobi theta functions are defined by
\begin{align*}
  \theta_{11}(z;\tau)
  & =
  \sum_{n \in \mathbb{Z}}
  q^{\frac{1}{2} \left( n+ \frac{1}{2} \right)^2} \,
  \E^{2 \pi \I \left(n+\frac{1}{2} \right) \,
    \left( z+\frac{1}{2} \right)
  }
  =
  \theta_1(z; \tau) ,
  \\[2mm]
%
  \theta_{10}(z;\tau)
  & =
  \sum_{n \in \mathbb{Z}}
  q^{\frac{1}{2} \left( n + \frac{1}{2} \right)^2} \,
  \E^{2 \pi \I \left( n+\frac{1}{2} \right) z}
  =
  \theta_2(z;\tau) ,
  \\[2mm]
%
  \theta_{00} (z;\tau)
  & =
  \sum_{n \in \mathbb{Z}}
  q^{\frac{1}{2} n^2} \,
  \E^{2 \pi \I  n  z}
  = \theta_3 (z;\tau) ,
  \\[2mm]
%
  \theta_{01} (z;\tau)
  & =
  \sum_{n \in \mathbb{Z}}
  q^{\frac{1}{2} n^2} \,
  \E^{2 \pi \I n \left( z+\frac{1}{2} \right) }
  =
  \theta_4(z;\tau) ,
\end{align*}
where we have also shown the relation to the conventional notations.


\end{document}